\documentclass[conference]{IEEEtran}
\usepackage{graphicx}
\usepackage[nobreak]{cite}
\usepackage{xurl}
\usepackage{xspace}
\usepackage{enumitem}
\usepackage[fleqn]{amsmath}
\usepackage{multicol}
\usepackage{newtxmath}

\makeatletter
  \let\MYcaption\@makecaption
\makeatother
\usepackage[font=footnotesize]{subcaption}
\makeatletter
  \let\@makecaption\MYcaption
\makeatother

\usepackage{framed}
\newcommand{\Conclusion}[1]{\begin{framed}\noindent #1\end{framed}}

\usepackage{cleveref}
\Crefname{figure}{Figure}{Figures}
\crefname{figure}{Fig.}{Figs.}
\crefname{table}{Table}{Tables}
\crefname{section}{Section}{Sections}

\newcommand{\Heading}[1]{\textbf{#1.}}
\newcommand{\RQ}[1]{\textit{RQ}${}_{\mathrm{#1}}$}

\newcommand{\Refactoring}[1]{\textsf{#1}}
\newcommand{\Code}[1]{\texttt{#1}}

\newcommand{\Cleft}{\mathit{Left}}
\newcommand{\Cmid}{\mathit{Mid}}
\newcommand{\Cright}{\mathit{Right}}

\newcommand{\Sim}{\mathit{Sim}}
\newcommand{\Size}[1]{\lvert#1\rvert}

\newcommand{\RM}{RMiner\xspace}

\begin{document}

\title{How Much Can a Behavior-Preserving Changeset Be Decomposed into Refactoring Operations?}

\author{%
\IEEEauthorblockN{Kota Someya}
  \IEEEauthorblockA{%
    \textit{School of Computing}\\
    \textit{Institute of Science Tokyo}\\
    Tokyo, Japan\\
    someya@se.comp.isct.ac.jp}
  \and
  \IEEEauthorblockN{Lei Chen}
  \IEEEauthorblockA{%
    \textit{School of Computing}\\
    \textit{Institute of Science Tokyo}\\
    Tokyo, Japan\\
    chenlei@se.comp.isct.ac.jp}
  \and
  \IEEEauthorblockN{Michael J. Decker}
  \IEEEauthorblockA{%
    \textit{Department of Computer Science}\\
    \textit{Bowling Green State University}\\
    Bowling Green, OH, USA\\
    mdecke@bgsu.edu}
  \and
  \IEEEauthorblockN{Shinpei Hayashi}
  \IEEEauthorblockA{%
    \textit{School of Computing}\\
    \textit{Institute of Science Tokyo}\\
    Tokyo, Japan\\
    hayashi@comp.isct.ac.jp}
}

\maketitle
\pagestyle{plain}
\thispagestyle{plain}

\begin{abstract}
Developers sometimes mix behavior-preserving modifications, such as refactorings, with behavior-altering modifications, such as feature additions.
Several approaches have been proposed to support understanding such modifications by separating them into those two parts.
Such refactoring-aware approaches are expected to be particularly effective when the behavior-preserving parts can be decomposed into a sequence of more primitive behavior-preserving operations, such as refactorings, but this has not been explored.
In this paper, as an initial validation, we quantify how much of the behavior-preserving modifications can be decomposed into refactoring operations using a dataset of functionally-equivalent method pairs.
As a result, when using an existing refactoring detector, only 33.9\% of the changes could be identified as refactoring operations.
In contrast, when including 67 newly defined functionally-equivalent operations, the coverage increased by over 128\%.
Further investigation into the remaining unexplained differences was conducted, suggesting improvement opportunities.
\end{abstract}

\section{Introduction}\label{s:introduction}

Refactoring is a common practice in software development to improve code readability/maintainability. 
It involves modifying the internal structure of code while preserving its external behavior \cite{Fowler2018}.
However, in real-world software development practices, developers may mix behavior-preserving modifications (e.g., refactoring) with behavior-altering modifications that introduce functional changes, i.e., \emph{floss refactoring} \cite{murphy-software2008,murphy-hill2012how-we}.
It may result in commits with \emph{tangled changes}, which hinder the review process \cite{tangled-changes,refactoring-aware-review-systematic-mapping}.

Addressing this, a series of authors propose refactoring-aware code review approaches\cite{ref-aware-review,rediffs,brito2021raid}.
These approaches facilitate a clearer understanding of changes during the review process by detecting fine-grained behavior-preserving operations using a refactoring detection tool so that the behavior-preserving modifications can be separated from behavior-altering modifications.
In essence, such refactoring-aware approaches are expected to be particularly effective when behavior-preserving parts can be extracted from a difference by using a sequence of behavior-preserving operations, including those detectable by refactoring detection tools, thus allowing developers to focus on the remaining behavior-altering parts.
However, in practice, the concept of behavioral equivalence includes a wide range of operations, from simple renamings to substantial internal structural changes such as algorithm substitution. 
Therefore, we argue that there are inherent limitations in fully explaining functionally equivalent parts using sequences of behavior-preserving operations.

As such, in this paper, we investigate the usefulness of a refactoring-aware code review approach.
Specifically, we focus on functionally equivalent code diffs and quantify how much of them can be decomposed into sequences of generalizable behavior-preserving operations, such as refactoring, and what remains unexplained.
In our evaluation, we do not address large-scale refactorings that affect the entire project. 
Instead, we focus on the limitations of difference decomposition at the method level, using a refactoring detection tool and sequences of behavior-preserving operations.

For the dataset, we utilize the Functionally Equivalent Method Pairs (FEMP) dataset \cite{femp}, which contains pairs of Java methods collected from open-source software (OSS) that are functionally equivalent. 
To evaluate the decomposition capability and detection limitations of existing refactoring tools, we use RefactoringMiner \cite{rminer, refminer2} as a representative. Results show that a refactoring detector (i.e., RefactoringMinor) can only be used to explain up to 34\% of the changes.
To provide a more comprehensive explanation of behavior-preserving changes, we introduce a catalog of 67 additional behavior-preserving operations and show that it can explain 77.5\% of changes, an increase of 128\%. 
Finally, we discuss the limitations of decomposing functionally equivalent code into behavior-preserving operations.

\def\RQone{To what extent can a refactoring detection tool decompose functionally equivalent changes using its existing techniques and catalogs?}
\def\RQtwo{To what extent can a refactoring detection tool decompose functionally equivalent changes when using an iterative approach?}
\def\RQthree{To what extent can functionally equivalent changes be decomposed via a more comprehensive set of behavior-preserving operations?}

More formally, we study the following research questions (RQs):
\begin{itemize}
  \item \textbf{\RQ{1}}: \RQone
  \item \textbf{\RQ{2}}: \RQtwo
  \item \textbf{\RQ{3}}: \RQthree
\end{itemize}

Answering these RQs results in the following contributions:
\begin{itemize}
  \item We demonstrate the decomposition capability of a refactoring detection tool applied as is and iteratively.
  \item We introduce an extended catalog of method-level behavior-preserving operations.
  \item We show that the extended catalog can be used to decompose code changes much more effectively into behavior-preserving operations.
\end{itemize}

The remainder of this paper is structured as follows.
In \cref{s:method}, we explain our methodology.
\Cref{s:empirical} explains the results of our empirical study.
Finally, \cref{s:conclusion} provides the conclusion and future work.

\section{Methodology}\label{s:method}

\begin{figure}[tb]\centering
  \includegraphics[width=\linewidth]{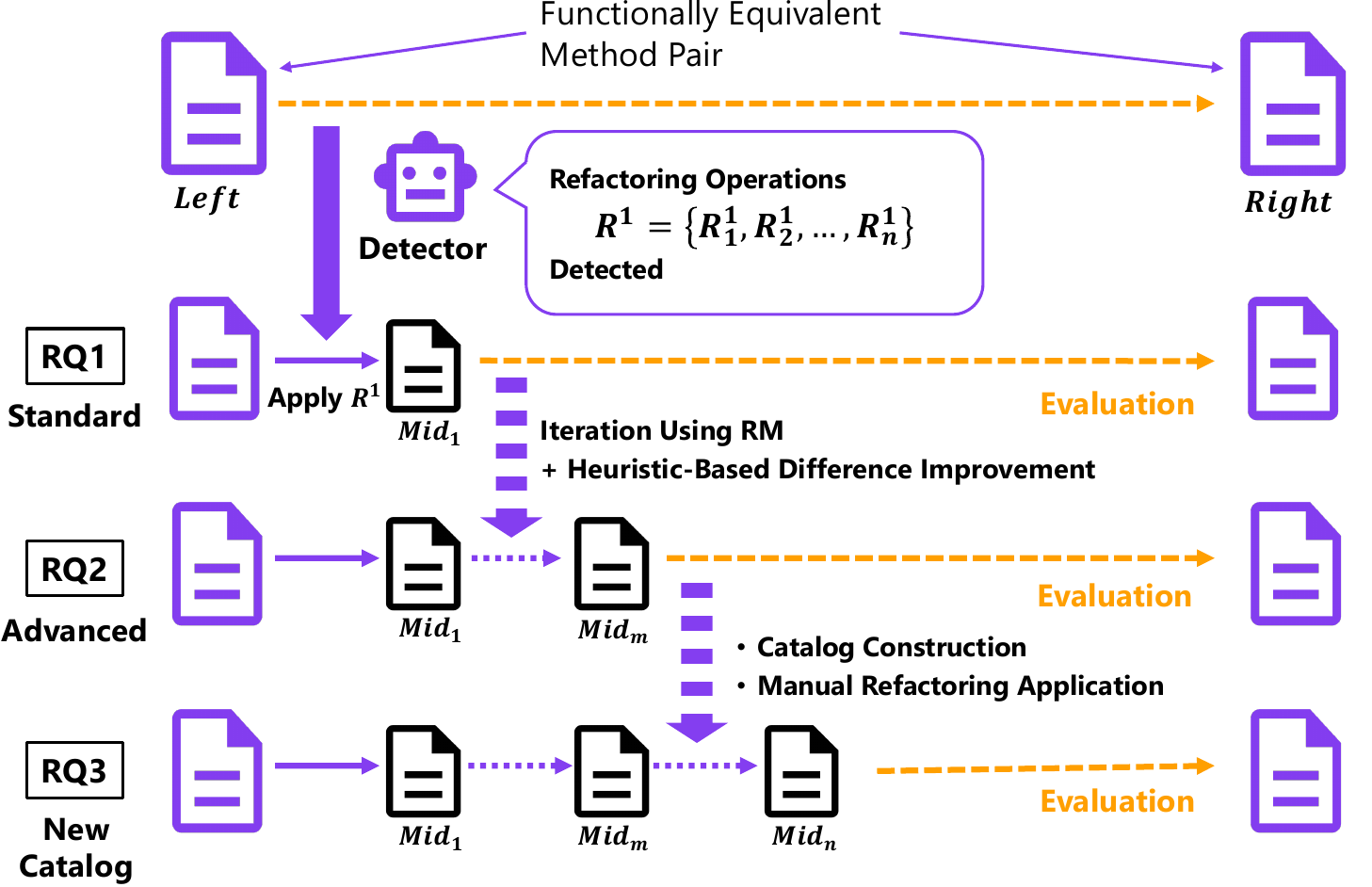}
  \caption{Overview of the study.}\label{fig:rq-flow}
\end{figure}

\Cref{fig:rq-flow} gives an overview of the methodology behind our investigation.
Our investigation involves three rounds of study.
For round one (i.e., \RQ{1}), we utilize as input a pair of functionally equivalent methods $\Cleft$ and $\Cright$.
We apply a refactoring detector and perform the detected operations on $\Cleft$, resulting in the intermediate code $\Cmid_1$.
We then evaluate the remaining differences between $\Cmid_1$ and $\Cright$.
For the second round (i.e., \RQ{2}), we rename the methods and parameters of $\Cmid_1$ to match those of $\Cright$, and then iteratively perform refactoring detection and application of those refactorings.
This represents a best-case scenario for current refactoring detection approaches.
We once again analyze the remaining changes, and we propose an extended catalog of behavior-preserving operations, which we apply in round three (i.e., \RQ{3}).
Finally, we investigate any remaining unexplained differences and provide improvement opportunities.

\subsection{Dataset}

In this paper, we use the FEMP dataset \cite{femp} to evaluate a refactoring detection tool and construct a more comprehensive catalog.  
The FEMP dataset consists of pairs of functionally equivalent Java methods collected from a large number of open-source projects. 
Here, \emph{functionally equivalent} means returning identical outputs for any given input, i.e., exhibiting the same external behavior.
Even if the internal algorithms or logic differ significantly, they are still considered functionally equivalent as long as their input-output behavior is identical.
We randomly selected 100 pairs from the FEMP dataset.

\subsection{Refactoring Detection Tool}

We use RefactoringMiner ver.\ 3.0.7  (\RM) \cite{refminer2} as a representative refactoring detector to conduct a detailed analysis of the capabilities of current refactoring detectors in explaining behavior-preserving changes.
We selected \RM because it has shown the highest accuracy among refactoring detection tools in previous studies \cite{ref-det-tools}.  
In addition, \RM supports 102 refactoring operations, including minor ones, making it considerably more comprehensive than other tools.
As such, cases where \RM fails to detect certain refactoring operations are likely to suggest issues commonly shared by existing detection techniques.

Only twelve method-level refactorings supported by \RM were detectable in our dataset.
However, we exclude two of them: \Refactoring{Add/Remove Parameter}, as applying them individually violates behavior preservation, making them false positives.
The ten refactorings used are listed in \cref{tab:ref-ops}.

\begin{table}[tb]\centering
  \caption{Refactorings Detectable by \RM}\label{tab:ref-ops}
  \footnotesize
  \begin{tabular}{c}\hline\vspace{-1.3em}\\
  \begin{minipage}{.95\linewidth}
  \begin{multicols}{2}
  \begin{itemize}[leftmargin=1em]
    \item \Refactoring{Add Method Annotation}
    \item \Refactoring{Add Variable Modifier}
    \item \Refactoring{Change Variable Type}
    \item \Refactoring{Extract Variable}
    \item \Refactoring{Inline Variable}
    \item \Refactoring{Remove Method Annotation}
    \item \Refactoring{Remove Variable Modifier}
    \item \Refactoring{Rename Method}
    \item \Refactoring{Rename Parameter}
    \item \Refactoring{Rename Variable}
  \end{itemize}
  \end{multicols}
  \end{minipage}\\\vspace{-0.5em}~\\\hline
  \end{tabular}
\end{table}

\subsection{Refactoring Applier}

To evaluate the decomposition capability of detected refactoring operations, we implemented a refactoring applier that automatically applies sequences of such operations.
The applier takes the output of \RM as input and applies them using the Eclipse JDT \cite{eclipse-jdt}.
\Cref{tab:ref-ops} also shows the refactoring coverage of the applier.
Of the 10 refactorings detected at the method level, 8 out of 10 refactorings can be automatically applied.
For \Refactoring{Extract Variable} and \Refactoring{Inline Variable}, we decided to apply them manually because the output of \RM does not include sufficient information to precisely perform these refactorings.

Note that in all cases, we verified that functional equivalence is preserved when applying each operation, regardless of whether it is applied automatically or manually, by running the test cases provided in the FEMP dataset.

\subsection{Evaluation Metric}

The following formula is used to quantify how well a set of behavior-preserving operations explains the differences between two functionally equivalent code fragments $\Cleft$ and $\Cright$, with $\Cmid$ representing the application of those operations to $\Cleft$:
\[
  \Sim(\Cmid, \Cleft, \Cright) = 1 - \frac{\Size{\Delta(\Cmid, \Cright)}}{\Size{\Delta(\Cleft, \Cright)}}.
\]
Here, $\Size{\Delta(c, c')}$ is the size of the delta between code fragments $c$ and $c'$, calculated as the total number of added and deleted tokens obtained using srcDiff~\cite{srcDiff}.
We adopt srcDiff, as it completely preserves all source code text, making it well-suited for quantifying the size of unexplained differences.
Conceptually, $\Sim$ computes the similarity based on the size of the delta between $\Cmid_k$ and $\Cright$, normalized by the original delta between $\Cleft$ and $\Cright$, to assess how much of the delta has been explained.
$\Sim$ ranges from 1, meaning 100\% of the changes between $\Cleft$ and $\Cright$ are explained by the behavior-preserving operations, to 0, meaning none of the changes can be explained.

\section{Empirical Study}\label{s:empirical}

The investigation results to answer the three RQs are summarized below.
All results, including the application process and intermediate results of operations, are available in the supplemental package \cite{artifact}.

\subsection{\RQ{1}: Standard Detector Usage}

\subsubsection{Motivation}

In this RQ, we investigate the decomposition capability of the refactoring detector in standard usage.

\subsubsection{Study Design}

First, we input a target method pair $(\Cleft, \Cright)$ into \RM and obtain a set of detected refactoring operations $R^1=\{R^1_1, R^1_2, \dots\}$. 
Next, we sequentially apply the applicable operations from $R^1$ to $\Cleft$ to obtain an intermediate code fragment $\Cmid_1$. 
Finally, we use $\Sim$ to measure how well $\Cmid_1$ explains the differences between $\Cleft$ and $\Cright$.

\begin{figure}[tb]\centering
  \begin{subfigure}[b]{\linewidth}\centering
    \includegraphics[width=\linewidth]{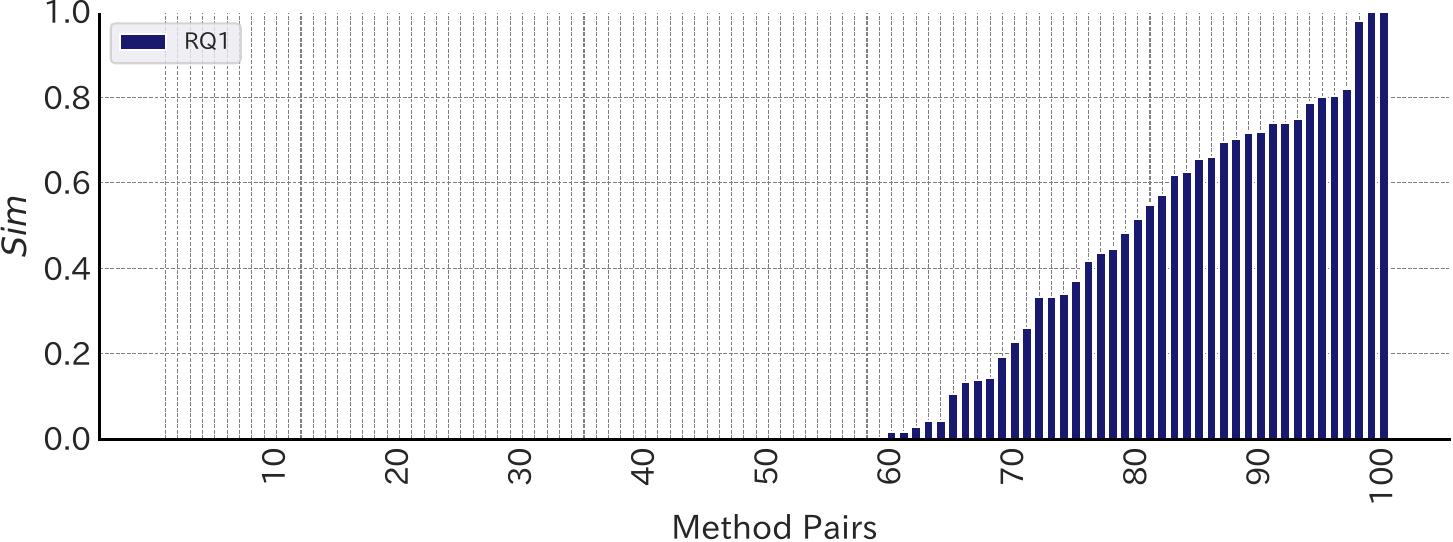}
    \caption{$\Sim$ results for \RQ{1}.}\label{f:rq1-distribution}
  \end{subfigure}
  \begin{subfigure}[b]{\linewidth}\centering
    \includegraphics[width=\linewidth]{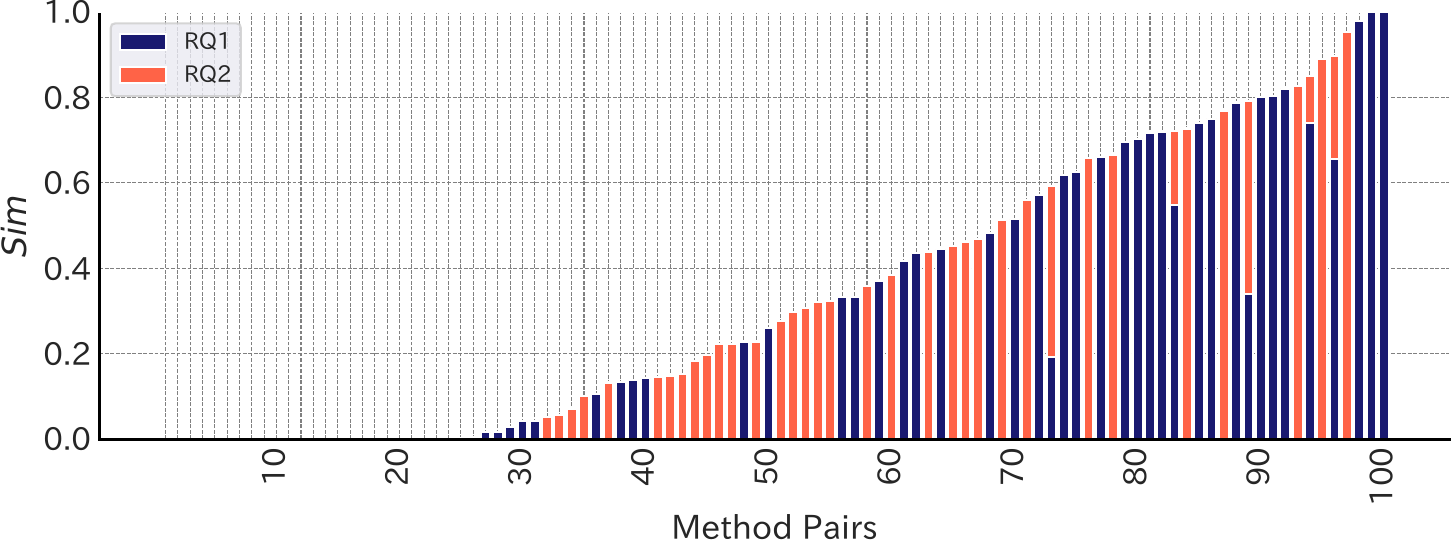}
    \caption{$\Sim$ results for \RQ{2}.}\label{f:rq2-distribution}
  \end{subfigure}
  \begin{subfigure}[b]{\linewidth}\centering
    \includegraphics[width=\linewidth]{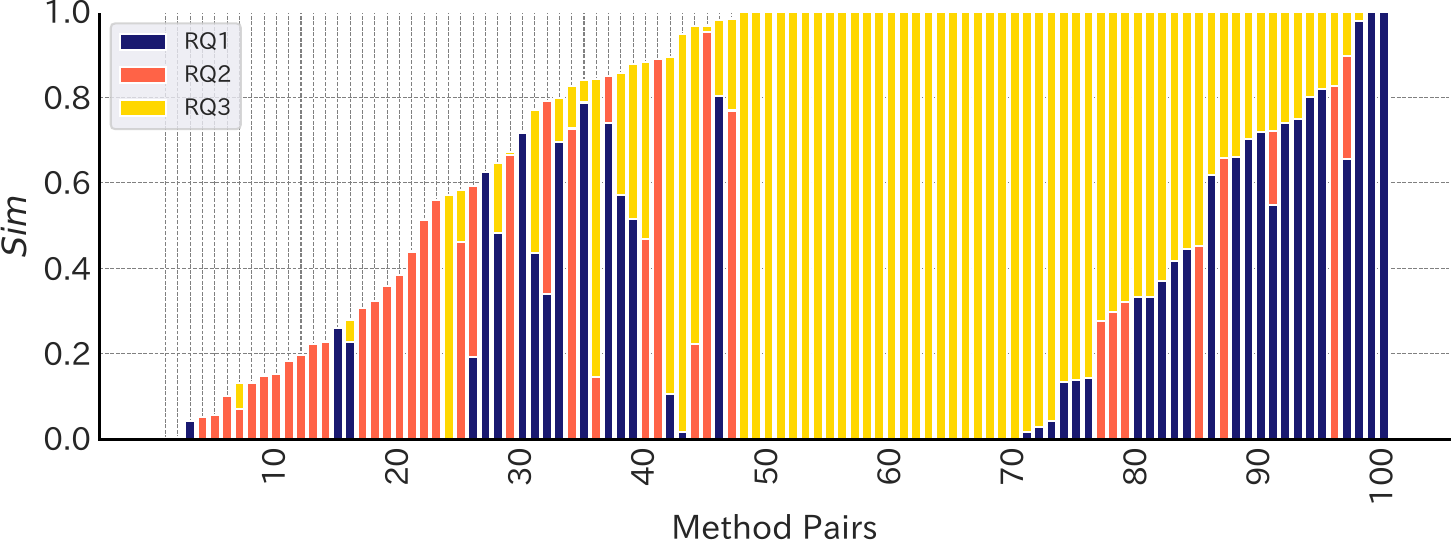}
    \caption{$\Sim$ results for \RQ{3}.}\label{f:rq3-distribution}
  \end{subfigure}
  \caption{Changes in $\Sim$ across all RQs.}\label{f:result}
\end{figure}

\subsubsection{Results}

The results are shown in \cref{f:rq1-distribution}. 
This figure presents the $\Sim$ values for each method pair, sorted in ascending order. 
42 of 100 pairs had at least one refactoring operation detected by \RM that could be applied.  
The average $\Sim$ across all method pairs is 0.2.
Only two pairs achieved a $\Sim$ of 1, indicating that all modifications were decomposed via a single round of detection and application.

\begin{table*}[tb]\centering
  \caption{Catalog of Behavior-Preserving Operations}\label{tab:catalog}
  \footnotesize
  \begin{tabular}{c}\hline\vspace{-1.3em}\\
  \begin{minipage}{.97\textwidth}
  \begin{multicols}{3}
  \begin{itemize}[leftmargin=1em] %
    \item \Refactoring{Apply Constant Folding}
    \item \Refactoring{Apply De Morgan's Law}
    \item \Refactoring{Apply Negation as Inequality}
    \item \Refactoring{Conditional to Expression}\cite{micro-change}
    \item \Refactoring{Consolidate Variable Declaration and Initialization}
    \item \Refactoring{Decompose Conditional Branch}
    \item \Refactoring{Factor Out Coefficient}
    \item \Refactoring{Factor Out Inequality as Negation}
    \item \Refactoring{Inline Return Variable}
    \item \Refactoring{Introduce Cast}
    \item \Refactoring{Introduce Constant to Comparison Expression}
    \item \Refactoring{Introduce Dead Code}
    \item \Refactoring{Introduce Parentheses}
    \item \Refactoring{Introduce Return Variable}
    \item \Refactoring{Merge Conditional Branch}
    \item \Refactoring{Merge Variable Declaration}
    \item \Refactoring{Move Statement Across Try}
    \item \Refactoring{Remove Branch by Pre-Assignment}
    \item \Refactoring{Remove Cast}
    \item \Refactoring{Remove Dead Code}\cite{Fowler2018}
    \item \Refactoring{Remove Double Negation}
    \item \Refactoring{Remove Parentheses}
    \item \Refactoring{Remove Unused Variable}
    \item \Refactoring{Replace Array Declaration Style}
    \item \Refactoring{Replace Assignment with Compound Assignment}
    \item \Refactoring{Replace Compound Assignment with Assignment}
    \item \Refactoring{Replace Conditional Operator with If}
    \item \Refactoring{Replace For with Foreach}
    \item \Refactoring{Replace Foreach with For}
    \item \Refactoring{Replace Guard Clause with Conditional}
    \item \Refactoring{Replace If with Switch}
    \item \Refactoring{Replace Inclusive Comparison with Exclusive}
    \item \Refactoring{Replace Nested Conditional with Guard Clauses}\cite{Fowler2018}
    \item \Refactoring{Replace Numeric Representation}
    \item \Refactoring{Replace Postfix Increment/Decrement with Prefix}
    \item \Refactoring{Replace Prefix Increment/Decrement with Postfix}
    \item \Refactoring{Replace Switch with If}
    \item \Refactoring{Reverse Comparison Operator}
    \item \Refactoring{Reverse Conditional}\cite{reverse-conditional}
    \item \Refactoring{Split Chained Assignment}
    \item \Refactoring{Split Conditional Branch}
    \item \Refactoring{Split Variable Declaration}
    \item \Refactoring{Split Variable Declaration and Initialization}
    \item \Refactoring{Swap Commutative Operands}
    \item \Refactoring{Swap Conditional Branches}
    \item \Refactoring{Transpose Equation}
    \item \Refactoring{Unwrap Statement from Block}\cite{micro-change}
    \item \Refactoring{Wrap Statement in Block}\cite{micro-change}
  \end{itemize}
  \end{multicols}
  \end{minipage}\\\vspace{-0.5em}~\\\hline
  \end{tabular}
\end{table*}

\subsection{\RQ{2}: Iterative Detector Usage}

\begin{figure}[tb]\centering
  \includegraphics[width=\linewidth]{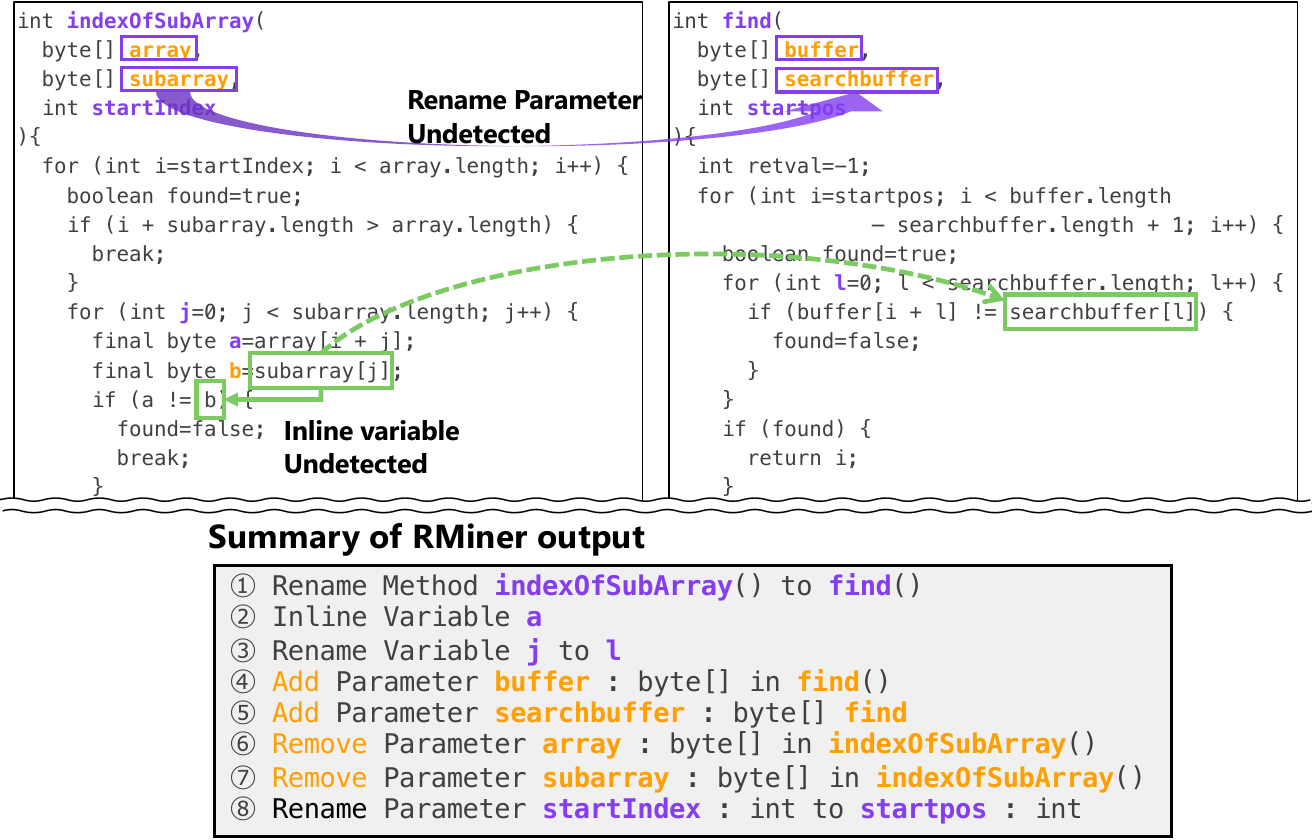}
  \caption{Insufficient detection by \RM (\#10998).}\label{fig:anti-rm-1}
\end{figure}

\subsubsection{Motivation}

\Cref{fig:anti-rm-1} shows the detection results by \RM for the method pair \#10998 in the FEMP dataset. 
In this example, only \Refactoring{Rename Method}, \Refactoring{Inline Variable}, \Refactoring{Rename Variable}, and \Refactoring{Rename Parameter} are correctly detected.
The changes to the first two parameters are incorrectly detected as a combination of \Refactoring{Add Parameter} and \Refactoring{Remove Parameter}. Moreover, we identified that issues caused by renaming are not limited to parameter renames. When parameter names and/or method names are different, especially when the internal structures of the code fragments vary significantly, the detector may fail to recognize the pair as corresponding methods, resulting in no detected operations. Thus, unifying the names can enable detection that would otherwise fail.

Furthermore, applying the detected operations to obtain a new version and then performing detection again can sometimes yield additional refactorings. 
For example, in the case mentioned above, reapplying \RM after applying the initially detected operations results in the correct detection of a previously missed \Refactoring{Inline Variable}. 
This result suggests that due to the characteristics of the detector's implementation, some refactoring operations may be missed or misinterpreted when the input difference is large. 
Reducing the size of the difference may improve the detector’s performance. 
In this RQ, we investigate the decomposition capability of the existing detection tool when manually applying rename refactorings and then iteratively using \RM.

\subsubsection{Study Design}

First, as discussed previously, because a detector may identify inapplicable refactorings (e.g., \Refactoring{Add/Remove Parameter}) or fail to detect any refactorings when the method/parameters are renamed, we manually detect and apply \Refactoring{Rename Method/Parameter} to unify the names.
The result of renaming is the intermediate code $\Cmid_2$.

After renaming to create $\Cmid_2$, we iteratively apply \RM. First, we apply \RM to $\Cmid_2$ to detect the sequence of operations between $(\Cmid_2, \Cright)$. 
The detected operations are then applied to $\Cmid_2$ to obtain $\Cmid_3$. 
This process is repeated until no refactoring operations are detected or no further operations can be applied. 
Specifically, for a given intermediate code fragment $\Cmid_t$, we use \RM to detect the operations  
$R^t=\{R^t_1, R^t_2, \dots\}$ between $(\Cmid_t, \Cright)$ and sequentially apply $R^t$ to $\Cmid_t$ to obtain $\Cmid_{t+1}$. 
This process is repeated until $R^t$ becomes an empty set or no further operations can be applied to $\Cmid_t$. 
The final code fragment obtained in this iterative process is denoted as $\Cmid_m$ and used as the target for evaluation.

\subsubsection{Results}

The results are shown in the same format as \RQ{1} in \cref{f:rq2-distribution}. 
76 of 100 method pairs had at least one refactoring operation that could be applied through the iterative detection process by \RM. 
As with \RQ{1}, only two method pairs achieved a $\Sim$ of 1, indicating that all differences were fully explained. 
The average $\Sim$ across the 100 pairs is 0.339, representing a marked increase from 20.0\% for \RQ{1}.

\subsection{\RQ{3}: Expansion of Behavior-Preserving Operations}

\begin{figure}[tb]\centering
  \includegraphics[width=\linewidth]{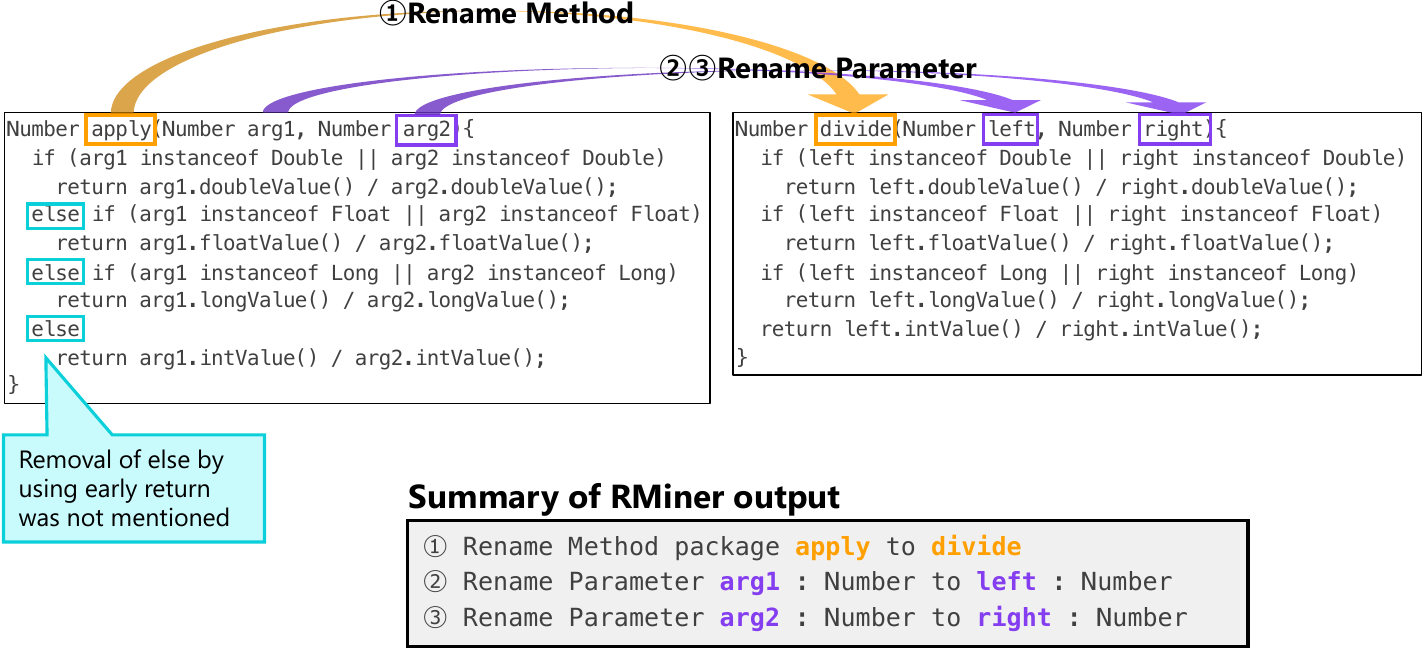}
  \caption{Example of an operation not implemented in \RM (\#11311).}\label{fig:anti-rm-2}
\end{figure}

\subsubsection{Motivation}

\Cref{fig:anti-rm-2} shows the detection results by \RM for the method pair \#11311 from the FEMP dataset.
In this example, an \Code{else} block is removed and replaced with a guard-clause.
However, this operation is not defined in \RM and thus remains undetected.
As demonstrated in \RQ{1} and \RQ{2}, one reason for the limited decomposition capability is the insufficient variety of refactoring operations.
To address this, we investigate the extent to which decomposition capability can be improved by introducing a new catalog with additional behavior-preserving operations.
We also examine the extent to which differences remain unexplained even after applying sequences of these operations.

\subsubsection{Study Design}

One of the authors created a catalog with behavior-preserving operations by observing the remaining differences between the intermediate code fragment $\Cmid_m$ and $\Cright$ (after applying the procedures for \RQ{1} and \RQ{2}).  
Based on these observations, we define additional behavior-preserving change operations.  
In constructing the catalog, we consider factors such as the frequency of occurrence, the feasibility of normalization, and the length of the code after normalization. 
Operations that are considered general-purpose based on the authors' judgment are selected.

For the code fragment $\Cmid_m$ obtained from \RQ{2}, we compare it to $\Cright$ and manually detect the additional behavior-preserving operations.
We iteratively apply these operations until no further operations from the extended catalog can be applied. This results in the final code fragment $\Cmid_k$.
Then, we calculate the $\Sim$ for $\Cmid_k$, $\Cleft$, and $\Cright$.

\subsubsection{Results}

\begin{figure}[tb]
  \centering
  \includegraphics[width=\linewidth]{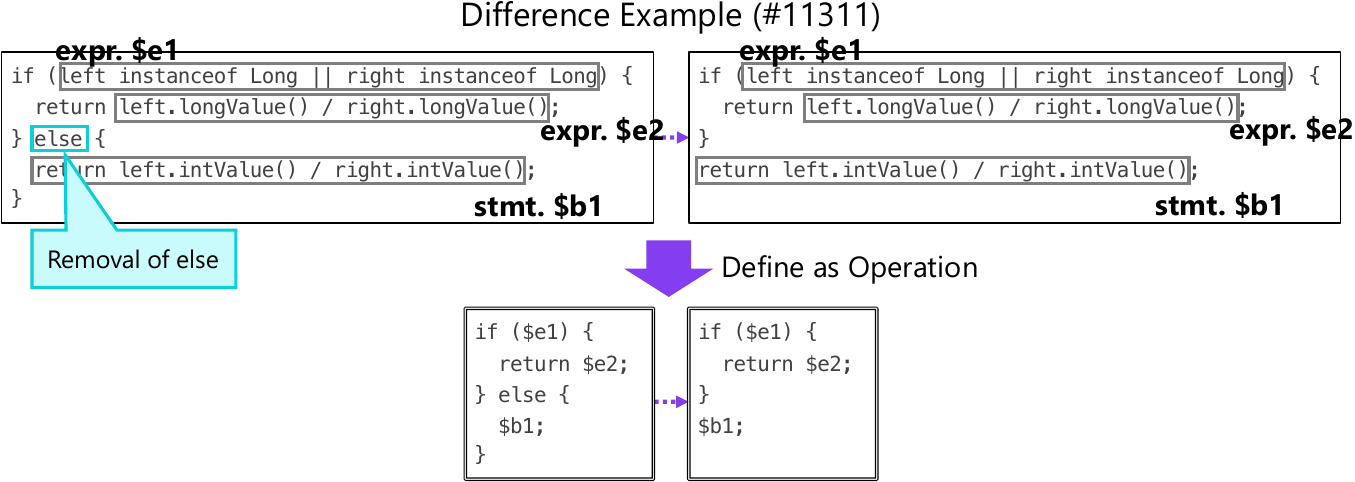}
  \caption{Extracting \Refactoring{Replace Nested Conditional with Guard Clauses}.}
  \label{fig:rq3-catalog}
\end{figure}

In the experiment conducted for \RQ{2}, 66.1\% of the differences remained unexplained.  
One of the authors carefully examined these remaining differences and constructed a catalog of 67 additional behavior-preserving operations that are not defined in \RM. 
Each operation in the catalog is structurally normalized based on transformations conforming to the Java language specification\cite{java-spec}.
\Cref{fig:rq3-catalog} shows the process of identifying the operation of \Refactoring{Replace Nested Conditional with Guard Clauses} from an actual example and normalizing it for the catalog.
Note that the same code fragment referenced (e.g., \Code{\$e1}) in the code before and after the modification is not required to be identical. 
Rather, it is expected that the code fragments are functionally equivalent. 

48 of the operations included in the catalog are listed in \cref{tab:catalog}. 
These are operations not defined in \RM, some of which may overlap with existing catalogs; we did our best to add references for those that have already been introduced in existing catalogs.
The remaining 19 operations not shown in the table are related to library usage, such as replacements between \Code{StringBuilder} and \Code{StringBuffer}.
These can be found in the complete catalog\cite{artifact}.

It should be noted that the operations defined here are behavior-preserving, but do not necessarily improve the readability/maintainability, e.g., \Refactoring{Introduce Dead Code}.
Therefore, we deliberately avoid calling them ``refactoring operations'' and instead refer to them as ``behavior-preserving operations.''
Also, these operations were directly extracted from our investigation of the dataset we examined and include well-known, representative changes.
We do not claim to have invented these changes, nor do we present this list as exhaustive.

The results are shown in \cref{f:rq3-distribution}, in the same format as in previous RQs.
Throughout the procedure, at least one operation was detected by \RM or through manual detection in all 100 method pairs. 
Furthermore, 53 of the 100 method pairs achieved a final $\Sim$ of 1, indicating that all modification differences were fully decomposed through the sequence of behavior-preserving operations. 
The average $\Sim$ for the 100 method pairs is 0.775.

\Conclusion{%
The average $\Sim$ in \RQ{1}, \RQ{2}, and \RQ{3} was 0.200, 0.339, and 0.775, respectively. 
These results indicate that introducing a catalog of behavior-preserving operations leads to an increase of 57.5\% over the single use of a refactoring detection tool, and an increase of 43.6\% over the iterative application of a detection tool.
This suggests that existing approaches to supporting change understanding can be further improved by refining detection tool usage and expanding the catalog. 
However, 22.5\% of the changes remained unexplained, indicating room for improvement.
}

\subsection{Threats to Validity}

\paragraph{Internal Validity}
The definition, manual detection, and application of operations in \RQ{3} involve subjective judgments solely by the first author, which may lead to a risk of misidentifying behavior-altering changes as behavior-preserving and/or overlooking operations that should have been detected.  
In both manual and automated applications, although we used the test cases provided by the dataset to ensure behavior preservation, there may still be untested boundary conditions that go undetected, resulting in incorrectly concluding functional equivalence.

\paragraph{External Validity}
Changes extracted from the FEMP dataset may be different from real-world projects.
Changes often occur at the class or project level, and a variety of programming languages may be involved.
As such, the generalizability of the results to other contexts remains uncertain.  

\section{Conclusion}\label{s:conclusion}

In this paper, we investigated the extent to which differences between functionally equivalent method pairs from the FEMP dataset can be decomposed using sequences of behavior-preserving operations.  
Specifically, we examined the decomposition capability of the refactoring detection tool RefactoringMiner under three conditions: standard usage (\RQ{1}), iterative usage (\RQ{2}), and with an expanded catalog of behavior-preserving operations (\RQ{3}).  
When applied to 100 method pairs, standard usage resulted in an average $\Sim$ of 0.200.
With the iterative application of detection and application, the $\Sim$ increased to 0.339.
Furthermore, by incorporating an extended catalog of additional behavior-preserving operations, the $\Sim$ reached 0.775.
Despite these improvements, 22.5\% of the differences remained that could not be fully decomposed through sequences of behavior-preserving operations.

Based on the obtained results, we discuss the following future directions:

\Heading{Further expansion of the catalog}
The current catalog remains incomplete. 
By analyzing a broader range of method pairs, we aim to expand its coverage and diversity. 
We also plan to include behavior-preserving operations beyond the method-level, such as those at the class and project levels.

\Heading{Automated detection and application}
In this study, the detection and application of newly defined operations were conducted manually.
To enable broader empirical studies and provide practical support, such as refactoring-aware code review, it is necessary to automate the identification and application of such operations from code differences.

\Heading{New approaches to change decomposition}
To promote understanding where the decomposition of differences proved challenging, it is important to explore alternative approaches instead of the decomposition of primitive operations.

\section*{Acknowledgments}
This work was supported in part by
  JSPS Grants-in-Aid for Scientific Research (JP23K24823, JP25K03102, JP25H01125, JP24H00692, and JP21KK0179) and
  US National Science Foundation: CNS 22-32593/32594.

\IEEEtriggeratref{10}
\bibliographystyle{IEEEtran}
\bibliography{references} 

\begin{thebibliography}{10}
\providecommand{\url}[1]{#1}
\csname url@samestyle\endcsname
\providecommand{\newblock}{\relax}
\providecommand{\bibinfo}[2]{#2}
\providecommand{\BIBentrySTDinterwordspacing}{\spaceskip=0pt\relax}
\providecommand{\BIBentryALTinterwordstretchfactor}{4}
\providecommand{\BIBentryALTinterwordspacing}{\spaceskip=\fontdimen2\font plus
\BIBentryALTinterwordstretchfactor\fontdimen3\font minus
  \fontdimen4\font\relax}
\providecommand{\BIBforeignlanguage}[2]{{%
\expandafter\ifx\csname l@#1\endcsname\relax
\typeout{** WARNING: IEEEtran.bst: No hyphenation pattern has been}%
\typeout{** loaded for the language `#1'. Using the pattern for}%
\typeout{** the default language instead.}%
\else
\language=\csname l@#1\endcsname
\fi
#2}}
\providecommand{\BIBdecl}{\relax}
\BIBdecl

\bibitem{Fowler2018}
M.~Fowler, \emph{Refactoring: Improving the Design of Existing Code},
  2nd~ed.\hskip 1em plus 0.5em minus 0.4em\relax Addison-Wesley, 2018.

\bibitem{murphy-software2008}
E.~Murphy-Hill and A.~P. Black, ``Refactoring tools: Fitness for purpose,''
  \emph{IEEE Software}, vol.~25, no.~5, pp. 38--44, 2008.

\bibitem{murphy-hill2012how-we}
E.~Murphy-Hill, C.~Parnin, and A.~P. Black, ``How we refactor, and how we know
  it,'' \emph{IEEE Transactions on Software Engineering}, vol.~38, no.~1, pp.
  5--18, 2012.

\bibitem{tangled-changes}
K.~Herzig and A.~Zeller, ``The impact of tangled code changes,'' in
  \emph{Proceedings of the 10th Working Conference on Mining Software
  Repositories (MSR 2013)}, 2013, pp. 121--130.

\bibitem{refactoring-aware-review-systematic-mapping}
F.~Coelho, T.~Massoni, and E.~L.G.~Alves, ``Refactoring-aware code review: A
  systematic mapping study,'' in \emph{Proceedings of the 3rd IEEE/ACM
  International Workshop on Refactoring (IWoR 2019)}, 2019, pp. 63--66.

\bibitem{ref-aware-review}
X.~Ge, S.~Sarkar, J.~Witschey, and E.~Murphy-Hill, ``Refactoring-aware code
  review,'' in \emph{Proceedings of the IEEE Symposium on Visual Languages and
  Human-Centric Computing (VL/HCC 2017)}, 2017, pp. 71--79.

\bibitem{rediffs}
S.~Hayashi, S.~Thangthumachit, and M.~Saeki, ``{REdiffs}: Refactoring-aware
  difference viewer for {Java},'' in \emph{Proceedings of the 20th Working
  Conference on Reverse Engineering (WCRE 2013)}, 2013, pp. 487--488.

\bibitem{brito2021raid}
R.~Brito and M.~T. Valente, ``{RAID}: {T}ool support for refactoring-aware code
  reviews,'' in \emph{Proceedings of the 29th IEEE/ACM International Conference
  on Program Comprehension (ICPC 2021)}, 2021, pp. 265--275.

\bibitem{femp}
Y.~Higo, ``Dataset of functionally equivalent {Java} methods and its
  application to evaluating clone detection tools,'' \emph{IEICE Transactions
  on Information and Systems}, vol. E107.D, no.~6, pp. 751--760, 2024.

\bibitem{rminer}
N.~Tsantalis, M.~Mansouri, L.~Eshkevari, D.~Mazinanian, and D.~Dig, ``Accurate
  and efficient refactoring detection in commit history,'' in \emph{Proceedings
  of the 40th IEEE/ACM International Conference on Software Engineering (ICSE
  2018)}, 2018, pp. 483--494.

\bibitem{refminer2}
N.~Tsantalis, A.~Ketkar, and D.~Dig, ``{RefactoringMiner} 2.0,'' \emph{IEEE
  Transactions on Software Engineering}, vol.~48, no.~3, pp. 930--950, 2022.

\bibitem{ref-det-tools}
L.~Tan and C.~Bockisch, ``A survey of refactoring detection tools,'' in
  \emph{Proceedings of the Workshops of the Software Engineering Conference,
  CEUR-WS}, vol. 2308, 2019, pp. 100--105.

\bibitem{eclipse-jdt}
{Eclipse Foundation}, ``Eclipse {Java} development tools ({JDT}),''
  \url{https://projects.eclipse.org/projects/eclipse.jdt}.

\bibitem{srcDiff}
M.~J. Decker, M.~L. Collard, L.~G. Volkert, and J.~I. Maletic, ``{srcDiff}: A
  syntactic differencing approach to improve the understandability of deltas,''
  \emph{Journal of Software: Evolution and Process}, vol.~32, no.~4, pp.
  e2226:1--31, 2020.

\bibitem{artifact}
\BIBentryALTinterwordspacing
K.~Someya, L.~Chen, M.~J. Decker, and S.~Hayashi, ``Artifact for ``{H}ow much
  can a behavior-preserving changeset be decomposed into refactoring
  operations?'','' Zenodo, 2025. [Online]. Available:
  \url{https://doi.org/10.5281/zenodo.16443373}
\BIBentrySTDinterwordspacing

\bibitem{micro-change}
L.~Chen, M.~Lanza, and S.~Hayashi, ``Understanding code change with
  micro-changes,'' in \emph{Proceedings of the 40th IEEE International
  Conference on Software Maintenance and Evolution (ICSME 2024)}, 2024, pp.
  363--374.

\bibitem{reverse-conditional}
B.~Murphy and M.~Fowler, ``{Reverse Conditional},''
  \url{https://www.refactoring.com/catalog/reverseConditional.html}, 2006.

\bibitem{java-spec}
J.~Gosling, B.~Joy, G.~Steele, G.~Bracha, A.~Buckley, and D.~Smith, ``The
  {Java} language specification, {Java SE} 11 edition,''
  \url{https://docs.oracle.com/javase/specs/jls/se11/jls11.pdf}, 2018.

\end{thebibliography}

\end{document}